\documentclass[10pt,conference]{IEEEtran}
\IEEEoverridecommandlockouts
\usepackage{authblk}
\usepackage{cite}
\usepackage{amsmath,amssymb,amsfonts,dsfont}
\usepackage{algorithm}
\usepackage[noend]{algorithmic}
\usepackage{graphicx}
\usepackage{textcomp}
\usepackage{xcolor}
\usepackage{bm}
\usepackage{booktabs}
\usepackage{multirow}
\usepackage{lipsum}
\usepackage{url}

\newtheorem{remark}{Remark}

\DeclareMathOperator*{\argmin}{\arg\min}

\newcommand{\mbbR}{\mathbb{R}} % real number
\newcommand{\bda}{\boldsymbol{\alpha}} 
\newcommand{\bdb}{\boldsymbol{\beta}}

\newcommand{\mcX}{\mathcal{X}}
\newcommand{\mcY}{\mathcal{Y}}

\newcommand{\mrmd}{\mathrm{d}}
\newcommand{\mrme}{\mathrm{e}}

\makeatletter
\renewcommand{\maketag@@@}[1]{\hbox{\m@th\normalsize\normalfont#1}}
\makeatother

\IEEEoverridecommandlockouts\IEEEpubid{\makebox[\columnwidth]{ 978-1-6654-3540-6/22~\copyright~2022 IEEE \hfill} \hspace{\columnsep}\makebox[\columnwidth]{ }}
\begin{document}

%\title{The Constrained Optimal Transport Model for Mismatch Capacity of AWGN Channels}
\title{An Optimal Transport Approach to \\$\ $ the Computation of the LM Rate
\thanks{The first two authors contributed equally to this work.}
\thanks{$\dag$ Corresponding authors.}
}

% \and
% \and
% \and
% \and
% \and

\author[1]{Wenhao Ye}
\author[2]{Huihui Wu}
\author[1]{Shitong Wu}
\author[2]{Yizhu Wang}
\author[3$\dag$]{Wenyi Zhang}
\author[1$\dag$]{Hao Wu}
\author[2]{Bo Bai}
%\author[$\flat$]{Author}
\affil[1]{Department of Mathematical Sciences, Tsinghua University, Beijing 100084, China}
\affil[2]{Theory Lab, Central Research Institute, 2012 Labs, Huawei Tech. Co. Ltd., Hong Kong SAR, China}
%Department of Computer Science and Engineering, \authorcr The Hong Kong University of Science and Technology, Hong Kong SAR, China}
\affil[3]{Department of Electronic Engineering and Information Science, 
\authorcr University of Science and Technology of China, Hefei, Anhui 230027, China 
\authorcr
Email: wenyizha@ustc.edu.cn, hwu@tsinghua.edu.cn}

%\author{
%\IEEEauthorblockA{Wenhao Ye\;
%\textit{Tsinghua University \& Huawei Technologies Co., Ltd., Beijing, China.}\; yewh20@mails.tsinghua.edu.cn}

%\IEEEauthorblockA{Zhennan Zhou\;
%\textit{Peking University, Beijing, China.}\; zhennan@bicmr.pku.edu.cn}

%\IEEEauthorblockA{Zhongyi Huang\;
%\textit{Tsinghua University, Beijing, China.}\; zhongyih@tsinghua.edu.cn}

%\IEEEauthorblockA{Bo Bai\;
%\textit{Huawei Technologies Co., Ltd., Hong Kong, China.}\; ee.bobbai@gmail.com}
%}

\maketitle

\begin{abstract}
% 我建议，不要在摘要中写出，或者凝练出缩写形式。只在正文中，第一次出现的时候，给出凝练出缩写形式。之后，一般用缩写。个别特别必要的情况，也可以不用缩写。

% Wireless network capacity can be regarded as the most important performance metric for wireless communication systems. 参考这句话，点出mismatch capacity的意义和价值。然后说明，关于它的计算（要讲一个原因），得到了人们的广泛关注。
%
Mismatch capacity characterizes the highest information rate for a channel under a prescribed decoding metric, and is thus a highly relevant fundamental performance metric when dealing with many practically important communication scenarios. Compared with the frequently used generalized mutual information (GMI), the LM rate has been known as a tighter lower bound of the mismatch capacity. The computation of the LM rate,\footnote{\textcolor{black}{To our best knowledge, the name LM rate first appeared in the reference \cite{Merhav1994On}. The capital letter LM seems to be the abbreviation of Lower bound on the Mismatch capacity.}} however, has been a difficult task, due to the fact that the LM rate involves a maximization over a function of the channel input, which becomes challenging as the input alphabet size grows, and direct numerical methods (e.g., interior point methods) suffer from intensive memory and computational resource requirements. Noting that the computation of the LM rate can also be formulated as an entropy-based optimization problem with constraints, in this work, we transform the task into an optimal transport (OT) problem with an extra constraint. This allows us to efficiently and accurately accomplish our task by using the well-known Sinkhorn algorithm. Indeed, only a few iterations are required for convergence, due to the fact that the formulated problem does not contain additional regularization terms. Moreover, we convert the extra constraint into a root-finding procedure for a one-dimensional monotonic function. Numerical experiments demonstrate the feasibility and efficiency of our OT approach to the computation of the LM rate.

\end{abstract}


\section{Introduction} \label{sec_introduction}

The Shannon capacity of a channel characterizes the ultimate transmission efficiency limit of reliable communication over a channel \cite{Shannon1948}. It has played a fundamental role in the research of communication theory and directed the development of communication systems for decades.

In many scenarios of practical interest, however, the perfect knowledge about a channel may not be available or may not be fully utilized for implementing the transceivers. Important examples include channels with uncertainty (like fading in wireless communication systems) \cite{2002Fading}, with non-ideal transceiver hardware \cite{zhang2011general}, or with constrained receiver structure \cite{salz1995com}. A commonly adopted practice of the receiver under such circumstances is to use a prescribed decoding metric, which may not be matched to the actual channel transition probability, and the mismatch capacity has been introduced to characterize the highest information rate under a prescribed decoding metric; see, e.g., \cite{1998Reliable} \cite{2020Information} and references therein.

Unfortunately, the mismatch capacity is still an open problem to date \cite{1995Channel}. So instead of it, researchers have been focusing on deriving its achievable lower bounds. The generalized mutual information (GMI) is a relatively simple lower bound \cite{1993Information}, which has found extensive applications in various setups; see, e.g., \cite{2002Fading} \cite{zhang2011general} \cite{2012Nearest}. However, the LM rate  \cite{Merhav1994On} is a tighter lower bound, by replacing the independent and identically distributed (i.i.d.) codebook ensemble for the GMI by constant-composition codebook ensemble.\footnote{The LM rate is \textcolor{black}{still not} the tightest lower bound of the mismatch capacity. There are several ways of improving the GMI and the LM rate by considering more structured codebook ensembles \cite{2020Information}. These are beyond the scope of this paper.
}

From the perspective of optimization, the primal forms of the GMI and the LM rate are highly similar, except that for the LM rate there is an additional constraint on the marginal distribution of the sought-for maximizing joint probability distribution over the input and output alphabets; see, e.g., \cite[Thm. 1]{2000Mismatched}. Consequently, the dual form of the GMI is easy to compute, since it can be written as a maximization over a real number \cite[Eqn. (12)]{2000Mismatched}, which is readily solved by a one-dimensional line search. On the other hand, the dual form of the LM rate further involves a maximization over a function of the channel input \cite[Eqn. (11)]{2000Mismatched}. Consequently the computation of the LM rate is much more challenging, and there is rare work on the numerical computation of the LM rate. Interior point methods (e.g., \cite{sdpt3,2010cvx}) can be applied for computing the LM rate, but it is typically memory intensive and computationally expensive. It is therefore desirable to develop effective computation models and efficient numerical algorithms for the LM rate.

In this paper, we formulate the LM rate computation as an optimal transport (OT) problem with an extra constraint, motivated by the similarity between these two optimization problems. 
Our contribution consists of three parts.
First, we propose an OT model with an extra constraint for the LM rate. 
Second, \textcolor{black}{we show that} this model can be solved directly using the well-known Sinkhorn algorithm \cite{2013sinkhorn}. 
Since our OT model is strongly convex, no additional regularization is required. 
This guarantees that we only need \textcolor{black}{several hundred of} Sinkhorn iterations to converge rather than tens of thousands of iterations for classical OT problems.
Last, we show that the extra constraint \textcolor{black}{in the OT problem formulation} can be efficiently handled by finding the root of a one-dimensional monotone function, and the feasibility of the extra constraint is guaranteed at each iteration. Numerical experiments show that for the case of QPSK and 16-QAM, our proposed algorithm is efficient and accurate. We also point out that our method is highly scalable and can be easily generalized to large-scale cases, such as 256-QAM.
%
%Last, we show that the extra constraint can be efficiently handled by finding the unique root of a one-dimensional monotone function, and the feasibility of the extra constraint is guaranteed at each iteration.
%
% Furthermore, we apply our algorithm for computing the LM rate of several code modulation schemes, such as QPSK and M-QAM. 
%
%Numerical experiments shows that for the case of QPSK and 16-QAM, our proposed algorithm is robust and accurate. 
%
%It should be pointed out our proposed algorithm needs less RAM and less complexity in each iteration, thus it is also able to be applied to the larger scale case, \thatis, 256-QAM, which is almost impossible for the interior-point method.
%

%\newpage

%
The remaining \textcolor{black}{part} of this paper is organized as follows. 
In Section \ref{sec_formulation_2}, after briefly reviewing the basic definition of the LM rate, we write it in the form of OT model with an extra constraint.
Next, we present the numerical methods for this problem, including the Sinkhorn algorithm and the treatment of the extra constraint in Section \ref{sec_Sinkhorn_3}. 
In Section \ref{sec_num_4}, the simulation results demonstrate the advantages of our approach. We finally conclude the paper in Section \ref{sec_conclu_5}.

\section{Problem Formulation} \label{sec_formulation_2}
\subsection{The LM Rate}
We consider a discrete memoryless communication channel model with transition law $W(y|x)$ over the (finite, discrete) channel input alphabet $\mcX= \{x_1,\cdots,x_M\}\subset\mbbR^2$ and the channel output alphabet $\mcY \subset\mbbR^2$.
Thus, given a probability measure $P_{X}$ on $\mcX$, we are able to define the joint probability distribution $P_{XY}$ on $\mcX\times\mcY$ and the output distribution $P_{Y}$ on $\mcY$ by the following:
\begin{align*}
    & P_{XY}(x_i,B)= W(B|x_i)P_{X}(x_i), ~ \forall B\subset\mcY, \\
    & P_{Y}(B) = \sum_{i=1}^{M} W(B|x_{i})P_{X}(x_i), ~ \forall B\subset\mcY.
\end{align*}
%
%Similarly, we define the output distribution $P_{Y}$ on $\mcY$ by:}
%
%\begin{equation}
%    P_{Y}(B) = \sum_{i=1}^{M} W(B|x_{i})P_{X}(x_i), ~ \forall B\subset\mcY.
%\end{equation}
%
This discrete memoryless channel is in fact a mapping from the input alphabet $\mcX$ to the output alphabet $\mcY$ with the channel transition law $W(y|x)$.
\textcolor{black}{For transmission of rate $R$, a block length-$n$ codebook $\mathcal{C}$ consists of $2^{nR}$ vectors $\mathbf{x}(m) = (x^{(1)}(m),\cdots,x^{(n)}(m))\in\mcX^{n}$. The encoder maps the message $m$ uniformly randomly selected from the set $\mathcal{M} = \{1,\cdots,2^{nR}\}$ to the corresponding codeword $\mathbf{x}(m)$. After receiving the output sequence $\mathbf{y} = (y^{(1)},\cdots,y^{(n)})$, the decoder forms the estimation on the message $m$ following the decoding rule
\begin{equation*}
    \hat{m} = \argmin_{j\in\mathcal{M}}\sum_{k=1}^{n}d(x^{(k)}(j),y^{(k)}),
\end{equation*}
where $d : \mcX\times\mcY\to\mbbR$ is  a non-negative function called the \textit{decoding metric}.}

%A mismatched decoder for the channel is based on the \textit{decoding metric} $d : \mcX\times\mcY\to\mbbR$, which is a non-negative function that measures the distance between the input and output alphabet.

\textcolor{black}{With the notations defined above, the LM rate, as one of the achievable lower bound of the mismatch capacity, is defined as:}
%% With the notation defined above, we are able to give the definition formulae of the LM rate \cite{2000Mismatched}, which is one of the famous achievable rates and also an important lower bound of the mismatch capacity, in the following:
%
%
%In this case, we manage to compute the LM rate, which is one of the famous achievable rates, in the following way \cite{2000Mismatched}.
%
%Denote $D(\cdot\|\cdot)$ as the relative entropy function \cite{1993Information}, we focus on the LM rate $I_{LM}(P_{X})$ with fixed distribution $P_{X}$,
%
\begin{subequations}\label{IM}
	\begin{align}
	I_{LM}(P_{X}) & := \min_{\textcolor{black}{\gamma\in\mathcal{P}(\mcX\times\mcY)}} D(\gamma\|P_{X}P_{Y}) \label{IM_obj} \\
	\textrm{s.t.}\quad \int_{y\in\mcY}\gamma(x_i,y) \mrmd y &= P_X(x_i),\ \forall x_i\in\mcX, \label{IM_Px} \\
	\sum_{x_i\in\mcX}\gamma(x_i,y) &= P_Y(y),\ \forall y\in\mcY, \label{IM_PY} \\
	\int_{y\in\mcY}\sum_{x_i\in\mcX}\gamma d(x_i,y) \mrmd y &
	\le \int_{y\in\mcY}\sum_{x_i\in\mcX}P_{XY}d(x_i,y) \mrmd y. \label{IM_ineq}
	\end{align}
\end{subequations}
Here $D(\cdot\|\cdot)$ denotes the relative entropy function and
\textcolor{black}{$\mathcal{P}(\mcX\times\mcY)$ denotes the set of all joint probability distributions on $\mcX\times\mcY$.}
%$\gamma(x,y)$ is a joint probability distribution on $\mcX\times\mcY$. 
%The \textit{decoding metric} $d : \mcX\times\mcY\to\mbbR$ in \eqref{IM_ineq} is a non-negative function that measures the distance between the input and output alphabet.

%\textcolor{black}{Given a decoding metric $d(x,y)$, the decoder is defined as the mapping $\xi_d : \mathcal{Y}\to\mathcal{X}$ that following the decoding rule
%\begin{equation*}
%    \xi_{d}(y) = \argmin_{x_{k}\in\mathcal{X}}d(x_k, y).
%\end{equation*}}

%

%in which $H\in\mbbR^{2\times 2}$ indicates the mismatch due to some channel problems. In practice, $H$ is usually a combination of rotation and scaling transformation, and its specific form can not be fully obtained. For brevity, we call $H$ the mismatched matrix.

%\noindent where $\gamma$ is the probability distribution on $\mcX\times\mcY$ and satisfies the marginal condition \eqref{IM_Px},\ref{IM_PY}) and the random-coding error probability inequality \eqref{IM_ineq}).
%
% Considering that the mismatched decoding problem may be caused by channel uncertainty when the decoder has an incorrect channel
% estimate, we then adopt the decoding metric $d(x,y) = -\log( W(y|H x))$ in this paper to represent this problem, where $H\in\mbbR^{2\times 2}$ represents the noise and rotation by incorrect channel estimation.
%

\subsection{The OT Problem}
In \eqref{IM}, we can see \textcolor{black}{that} the LM rate is similar to the OT problem \cite{peyre2019computational} in terms of optimization objective and constraints. Naturally, we would like to derive an equivalent OT-type problem to \eqref{IM}, which creates a prerequisite for computing the LM rate by using the Sinkhorn algorithm \cite{2013sinkhorn}.

Consider a discrete probability distribution $\mu_i=P_X(x_i)$ on $\mcX$. It satisfies the following normalization and \textcolor{black}{average} power constraints:
%
%For the discrete memoryless channel, the input alphabets is a infinite set $\mcX = \{x_1,\cdots,x_M\}$. 
%
%Thus the input distribution $P_X$ on $\mcX$ is a discrete distribution $P_{X}(x_i) = \mu_i$. 
%
%Hence, we have the normalization as well as the power constraint:
%
\begin{equation}\label{pc}
    \sum_{i=1}^{M}\mu_{i} = 1,\quad\sum_{i=1}^{M}\mu_{i}\|x_{i}\|_{2}^{2} = 1.
\end{equation}
\textcolor{black}{For given channel transition law $W(y|x)$ and decoding metric $d(x,y)$, the right-hand side of \eqref{IM_ineq} is a constant:}
%
%In this way, by the definition of the distribution $P_{X Y}(x,y)$ and the decoding metric $d(x,y)=W(y|Hx)$, we are able to show the right-hand side of \eqref{IM_ineq}) is a constant, denoted by
%
\begin{equation}\label{const_T}
\begin{aligned}
    T&=\int_{\mcY}\sum_{i=1}^{M} P_{XY}(x_i,y)d(x_i,y) \mrmd y \\
    &= \int_{\mcY}\sum_{i=1}^{M} \mu_{i}W(y|x_i)d(x_i,y) \mrmd y.
\end{aligned}
\end{equation}
%
%and the constant $T$ is irrelevant with the parameter $\gamma$.
%
Moreover, it is easy to show that
\begin{align*}
    & D(\gamma\|P_{X}P_{Y})=\int_{\mcY}\sum_{i=1}^{M}
        \gamma(x_{i},y)\log\gamma(x_{i},y)\mrmd y-C, \\
    & C=\sum_{i=1}^MP_{X}(x_i)\log P_{X}(x_i)
        +\int_{\mcY}P_{Y}(y)\log P_{Y}(y) \mrmd y.
\end{align*}
% 即最大化D()的本质上是极小化熵
%
By neglecting the constant $C$, minimizing the objective function $D(\gamma\|P_{X}P_{Y})$ is equivalent to maximizing the entropy
%the objective function $D(\gamma\|P_{X}P_{Y})$ in \eqref{IM_obj}) can be replaced with the entropy function:
%
\begin{equation*}
    -\int_{\mcY}\sum_{i=1}^{M}\gamma(x_{i},y)
        \log\gamma(x_{i},y)\mrmd y.
\end{equation*}
%
%where the relative entropy function \cite{bai2020information} $D(\gamma\|P_{X}P_{Y})$ can be written as the sum of the entropy function \eqref{en_obj}) and the constant $A_1$ and $A_2$ that are irrelevant with the parameter $\gamma$:
%
%\begin{equation*}
%\begin{aligned}
%    D(\gamma\|P_{X}P_{Y})&=\int_{\mcY}\sum_{i=1}^{M}\gamma(x_{i},y)\log\gamma(x_{i},y)\mrmd y+A_{1}+A_{2},\\
%    A_{1} &= \int_{\mcX\times\mcY}\gamma(x,y)\log P_{X}(x)\mrmd x \mrmd y, \\
%    A_{2} &= \int_{\mcX\times\mcY}\gamma(x,y)\log P_{Y}(y) \mrmd x \mrmd y.
%\end{aligned}
%\end{equation*}
%
Thus, we have obtained the following optimal transport problem \eqref{LM_obj}-\eqref{LM_y} with an extra constraint \eqref{LM_ineq}:
%Above all, the optimization problem of solving $I_{LM}(P_{X})$ can be equivalently transformed into the formula below:
%
\begin{small}
\begin{subequations} \label{LM}
	\begin{align} 
	\min_{\textcolor{black}{\gamma\in\mathcal{P}(\mcX\times\mcY)}}&\int_{\mcY}\sum_{i=1}^{M}\gamma(x_{i},y)\log\gamma(x_{i},y)\mrmd y \label{LM_obj} \\
	\textbf{s.t.}~&\int_{\mcY}\gamma(x_{i},y) \mrmd y = \mu_i,\quad i = 1,\cdots,M, \label{LM_x} \\
	&\sum_{i=1}^{M}\gamma(x_{i},y) = P_{Y}(y),\quad \forall y\in \mcY, \label{LM_y} \\
	&\int_{\mcY}\sum_{i=1}^{M}\gamma(x_{i},y) d(x_i,y)\mrmd y \leq T. \label{LM_ineq}
	\end{align}
\end{subequations}
\end{small}
%
%where $T$ is a constant determined in equation \eqref{const_T}).
%
\begin{remark}
    %We admit that it is not rigorous to consider $\log\gamma(x_{i},y)$ in \eqref{LM_obj} as the cost function. However, from the achievable rate point of view, the original problem (2.92) in \cite{2020Information} is more close to the OT form. After some complicated and lengthy inequality scaling of the original problem, we can obtain the LM rate \eqref{LM_obj}-\eqref{LM_ineq}, and it is still very similar to the OT form. For detailed derivation, the interested readers are referred to Chapter 2 in \cite{csiszar2011information}. On the other hand, from the numerical point of view, the above optimization problem \eqref{LM_obj}-\eqref{LM_ineq} can be efficiently solved by the Sinkhorn algorithm. In these senses, we might as well call it an OT problem.
    \textcolor{black}{In the classical OT theory, the cost function is only related to distance and should be independent of variables. Thus, \eqref{LM_obj}-\eqref{LM_ineq} is not a standard OT form, since the ``cost function'' $\log\gamma(x_i,y)$ in \eqref{LM_obj} is related to the variables $\gamma(x_i,y)$. However, the optimization objective function \eqref{LM_obj} is obtained by complicated and lengthy inequality scaling of the original problem (2.29) in \cite{2020Information}, which is in standard OT form. For detailed derivation, the interested readers are referred to Chapter 2 in \cite{csiszar2011information}. On the other hand, from the numerical point of view, the optimization problem \eqref{LM_obj}-\eqref{LM_ineq} can be efficiently solved by the Sinkhorn algorithm, which is well-known for its high efficiency in solving OT problems. In these senses, we might as well call \eqref{LM_obj}-\eqref{LM_ineq} an OT problem.}
    
    % The OT form in \eqref{LM_obj}-\eqref{LM_ineq}, which is equivalent to the LM rate, is obtained after some inequality scaling. The whole derivation is too complicated and lengthy to reproduce here. 
    % First, from the numerical point of view, the above optimization problem can be efficiently solved by the Sinkhorn algorithm. Secondly, from the achievable rate point of view, the original problem (2.92) in \cite{2020information} is more close to the OT form. The OT form in \eqref{LM_obj}-\eqref{LM_ineq}, which is equivalent to the LM rate, is obtained after some inequality scaling. The whole derivation is too complicated and lengthy to reproduce here. The interesting readers is referred to Chapter 2 in \cite{csiszar2011information}.
    %The entropy-based optimization problem \eqref{IM}) and as well as its simplified form \eqref{LM}) are an OT problem. In fact, the relative entropy function in \eqref{IM_obj}) can be viewed as the general cost function in the OT problem. As it is shown in the paper \cite{bai2020information}, the cost function to measure the error probability can be controlled by the relative entropy function in \eqref{IM_obj}), thus we are able to view this as a general cost function.
\end{remark}

\section{THE Sinkhorn ALGORITHM} \label{sec_Sinkhorn_3}
In this section, we turn to the Sinkhorn algorithm for the equivalent OT problem to the LM rate. First, we need to discretize the integral in \eqref{LM}. We might as well consider a set of uniform grid points $\{y_j\}_{j=1}^N$ in $\mcY$, and the corresponding rectangular discretization of the integral. This leads to the discrete form of \eqref{LM}:
%In this section, we focus on the LM rate of the AWGN channels, and then adopt the well-known Sinkhorn algorithm to solve the optimization problem aforementioned.
%
%Firstly, we directly give the discrete form of \eqref{LM} as:
%
%
\begin{subequations}\label{OT}
\begin{align}
    \min_{\textcolor{black}{\gamma_{ij}}}\quad&\sum_{i=1}^{M}\sum_{j=1}^{N}\gamma_{ij}\log \gamma_{ij} \label{OT_obj} \\
    \textbf{s.t.}\quad&\sum_{j=1}^{N}\gamma_{ij} = \mu_i ,\quad i = 1,\cdots,M, \label{OT_Px} \\
    &\sum_{i=1}^{M}\gamma_{ij} = \nu_j,\quad j = 1,\cdots,N, \label{OT_Py} \\
    & \sum_{i=1}^{M}\sum_{j=1}^{N} d_{ij}\gamma_{ij}\le T. \label{OT_ineq}
\end{align}
\end{subequations}
Here $\gamma_{ij}=\gamma(x_i,y_j),\; \nu_j=P_Y(y_j)$ and $d_{ij}=d(x_i,y_j)$. 
%Thus, we have obtained the above discrete OT form.

\begin{remark}
   In this work, we illustrate our key idea through the rectangular \textcolor{black}{formula} with only first-order accuracy for numerical integration. The idea can be easily generalized to a higher-order numerical discretization, e.g., the trapezoid formula. \textcolor{black}{Not only that, the Sinkhorn algorithm, which we will discuss later in this section, only requires minor modifications to be suitable for higher-order discretization.}
   %Not only that, the Sinkhorn algorithm we will discuss later in this part is not affected by high-order discretization.
   %In this work, we illustrate our key idea through the rectangular formula with only first-order accuracy for numerical integration. This may result in low accuracy of the LM rate. However, this does not bore us for two reasons. First, we can easily generalize to a higher-order numerical discretization, e.g., the trapezoid formula. Secondly, the discrete OT form and corresponding Sinkhorn algorithm are not affected by the high-order discretization.
\end{remark}

Introducing the dual variables $\bda\in\mbbR^{M}, \bdb\in\mbbR^{N}$ and $\lambda\in\mbbR^{+}$, the Lagrangian of \eqref{OT} \textcolor{black}{can be written as}:
\begin{equation}
\begin{aligned}
	&\mathcal{L}(\gamma; \bda, \bdb, \lambda) = \sum_{i=1}^{M}\sum_{j=1}^{N}\gamma_{ij}\log\gamma_{ij} +\sum_{i=1}^{M}\alpha_{i}(\sum_{j=1}^{N}\gamma_{ij} - \mu_i)\\
	& + \sum_{j=1}^{N}\beta_{j}(\sum_{i=1}^{M}\gamma_{ij} - \nu_j)+\lambda(\sum_{i=1}^{M}\sum_{j=1}^{N}d_{ij}\gamma_{ij} - T).
\end{aligned}
\end{equation}
%
%
% Note that \eqref{OT_obj} is strictly convex.
%
Taking the derivative of $\mathcal{L}(\gamma; \bda, \bdb, \lambda)$ with respect to $\gamma_{ij}$ \textcolor{black}{leads to}
\begin{equation*}
    \gamma_{ij} = \phi_{i}\Lambda_{ij}\psi_{j},
\end{equation*}
in which
\begin{equation*}
    \phi_{i} = \mrme^{-\alpha_{i}-1/2},\;
    \psi_{j} = \mrme^{-\beta_{j}-1/2},\;
    \Lambda_{ij} = \mrme^{-\lambda d_{ij}}.
\end{equation*}
Substituting the above formula into \eqref{OT_Px} and \eqref{OT_Py} yields
\begin{equation*}
\begin{aligned}
&\phi_{i}\sum_{j=1}^{N}\Lambda_{ij}\psi_{j} = \mu_i,\quad  i = 1,\cdots,M,   \\
&\psi_{j}\sum_{i=1}^{M}\Lambda_{ij}\phi_{i} = \nu_j,\quad  j = 1,\cdots,N.  
\end{aligned}
\end{equation*}
Since $\Lambda_{ij}>0$, we can alternatively update $\phi_i$ and $\psi_j$ as follows:
%
%Then  the Sinkhorn’s iteration can be
%applied to update vectors $\bdphi=(\phi_{i})$ and $\bdpsi=(\psi_{j})$ and by pointwise computation
%
\begin{subequations}\label{sink_iter}
\begin{align}
&\phi_{i}^{(\ell + 1)} = \mu_{i}/\sum_{j=1}^{N}\Lambda_{ij}^{(\ell)}\psi_{j}^{(\ell)},\quad i = 1,\cdots,M, \label{sink_px} \\
&\psi_{j}^{(\ell + 1)} = \nu_{j}/\sum_{i=1}^{M}\Lambda_{ij}^{(\ell)}\phi_{i}^{(\ell+1)},\quad j = 1,\cdots,N. \label{sink_py}
\end{align}
\end{subequations}
This iterative formula is the well-known Sinkhorn algorithm \cite{sinkhorn1967diagonal}. Different from the classical OT problem, we need to deal with the extra constraint \eqref{OT_ineq}. By taking the derivative of $\mathcal{L}(\gamma; \bda, \bdb, \lambda)$ with respect to $\lambda$, we have
%
%After updating $\bdphi$ and $\bdpsi$ by the Sinkhorn algorithm, the dual variable $\lambda$ is then updated by solving the root of $G(\lambda)$, here the function $G(\lambda)$ is defined as:
%
\begin{equation}\label{G_lambda}
    G(\lambda)\triangleq \sum_{i=1}^{M}\sum_{j=1}^{N} \phi_{i}\psi_{j} d_{ij} \mrme^{-\lambda d_{ij}} - T=0.
\end{equation}
Thus, we can update $\lambda$ by finding the roots of
\begin{equation*}
    G(\lambda^{(\ell+1)})\triangleq \sum_{i=1}^{M}\sum_{j=1}^{N} \phi_{i}^{(\ell+1)}\psi_{j}^{(\ell+1)} d_{ij} \mrme^{-\lambda^{(\ell+1)} d_{ij}} - T.
\end{equation*}
\textcolor{black}{Noticing} that
\begin{equation*}
    G'(\lambda)= -\sum_{i=1}^{M}\sum_{j=1}^{N}\phi_{i}\psi_{j}d_{ij}^{2}\mrme^{-\lambda d_{ij}} < 0,
\end{equation*}
the function $G(\lambda)$ is monotonic. Thus, The problem of finding the root of $G(\lambda)$ can be easily solved by the Newton's method. The pseudo-code \textcolor{black}{of the proposed algorithm} is present in Algorithm \ref{alg:OT}.

% Further, we design the Sinkhorn algorithm for solving \eqref{OT} by cyclically updating the variable $\bdphi$, $\bdpsi$ and $\lambda$ at each iteration. Indeed, this is also called the block coordinate descent method.
%

\begin{algorithm}[H]
	\caption{The Sinkhorn Algorithm}
	\label{alg:OT}
	\begin{algorithmic}[1]
		\REQUIRE Decoding metric $d_{ij}$; Marginal distributions $\mu_i, \nu_j$; Iteration number $K$.
		%Error bound $err$.
		\ENSURE Minimal value of $\sum_{i=1}^{M}\sum_{j=1}^{N}\gamma_{ij}\log\gamma_{ij}$.
		\STATE \textbf{Initialization:} $\bm{\phi} = \mathbf{1}_{M}, \bm{\psi} = \mathbf{1}_{N}, \lambda = 1$;
		\FOR{$\ell = 1 : K$}
		%\WHILE{$\ell < K$}
		\STATE $\Lambda_{ij} \gets \mrme^{-\lambda d_{ij}}$,\quad $i = 1,\cdots,M,\ j = 1,\cdots,N$
		\FOR{$i = 1 : M$}
		\STATE $\phi_{i} \gets \mu_{i}/\sum_{j=1}^{N}\Lambda_{ij}\psi_{j}$
		\ENDFOR
		\FOR{$j = 1 : N$}
		\STATE $\psi_{j} \gets \nu_{j}/\sum_{i=1}^{M}\Lambda_{ij}\phi_{i}$
		\ENDFOR
		%\STATE Set $G(\lambda) = \sum_{i=1}^{M}\sum_{j=1}^{N} \alpha_{i}^{(\ell+1)}\beta_{j}^{(\ell+1)}\langle Hx_{i},y_{j}\rangle\Lambda_{ij}^{(\ell)}$
		\STATE Solve $G(\lambda) = 0$ for $\lambda$ with Newton's method
		%\ENDWHILE
		\ENDFOR
		\RETURN $\sum_{i=1}^{M}\sum_{j=1}^{N} \phi_{i}\psi_{j}\Lambda_{ij}\log\left(\phi_{i}\psi_{j} \Lambda_{ij}\right)$
	\end{algorithmic}
\end{algorithm}
Finally, we need to discuss the feasibility of the Sinkhorn iteration, especially \textcolor{black}{under} the extra constraint \eqref{OT_ineq}. Depending on the value of $G(0)$, there are two cases:
\begin{itemize}
    \item $G(0)>0$: In this case, $G(\lambda)=0$ has a unique solution on $(0,+\infty)$ since $G'(\lambda)<0$. Thus the extra constraint \eqref{OT_ineq} is obviously satisfied.
    
    \item $G(0)\le 0$: In this case, the extra constraint \eqref{OT_ineq} is already satisfied. We only need to set $\lambda=0$ instead of solving $G(\lambda)=0$ at  line 8 of Algorithm \ref{alg:OT}.
\end{itemize}

\section{NUMERICAL SIMULATIONS} \label{sec_num_4}
In this section, we use the model and algorithm developed in \textcolor{black}{previous sections} to \textcolor{black}{compute} the LM rate for different modulation schemes over the additive white Gaussian noise (AWGN) channels \textcolor{black}{subject to rotation and scaling} with the following transition law 
%
%In this section, we will carry out our proposed Sinkhorn algorithm over the additive white Gaussian noise (AWGN) channels, for the QPSK modulation scheme and the 16-QAM modulation scheme.
%
% Especially, we consider the mismatched decoding problem with the transition law
%
\begin{equation}\label{awgneq}
    Y = H X + Z, \ \text{with } Z\sim \mathcal{N}(0,\sigma_{Z}^{2}).
\end{equation}
The matrix $H\in\mbbR^{2\times 2}$ is a combination of rotation and scaling transformations as
\textcolor{black}{
\begin{equation*}
    H = \begin{pmatrix} \eta_1 & 0 \\ 0 & \eta_2 \end{pmatrix}\begin{pmatrix} \cos \theta & \sin \theta\\ -\sin \theta & \cos \theta \end{pmatrix}.
\end{equation*}}
Here the parameters $\eta_1,\;\eta_2$ indicate the scaling of the signal, and the parameter $\theta$ indicates the degree of rotation on the signal. In this work, we consider the case of $\eta_1=1$ and $\eta_2=\eta$. \textcolor{black}{Note that}, the specific values of $\eta$ and $\theta$ can not be priorly known.

%and its specific form can not be fully obtained due to some channel problems. 
%
The decoding metric is $d(x,y) := \|y-\hat{H}x\|_{2}^{2}$, where $\hat{H}$ is an approximation that is based on part of  the information we know about $H$. We consider the scenario in which the decoder is unaware of the mismatch effect. Thus, in the decoding rule, \textcolor{black}{we have} $\hat{H}=I$.
%performs a channel estimation with known information of $H$.
%
%We consider the scenario in the decoder is unaware of the mismatch effect, and therefore we take $\hat{H}=I$ in our decoding rule.
%
%Specifically, in each example we will calculate the LM rate under different mismatched cases, where $H$ has the form
%
%\begin{equation}
%    H = \begin{pmatrix} 1 & 0 \\ 0 & \eta \end{pmatrix} \cdot \begin{pmatrix} \cos \theta & \sin \theta\\ -\sin \theta & \cos \theta \end{pmatrix}.
%\end{equation}
%
%Here the parameter $\eta,\theta$ indicates the scaling and the degree of rotation on the signal, which results in the mismatch problem. 
%
%In this case, by directly integrating the Gaussian function,
%\stwu{\eqref{OT_ineq} can be written as
%\begin{equation}
%    \sum_{i=1}^{M}\sum_{j=1}^{N}\langle x_{i},y_{j}\rangle\gamma_{ij} \geq \sum_{i=1}^{M}\mu_{i}\langle H x_{i},x_{i}\rangle.
%\end{equation}}

\begin{figure}[H]
	\centerline{\includegraphics[width=0.5\textwidth]{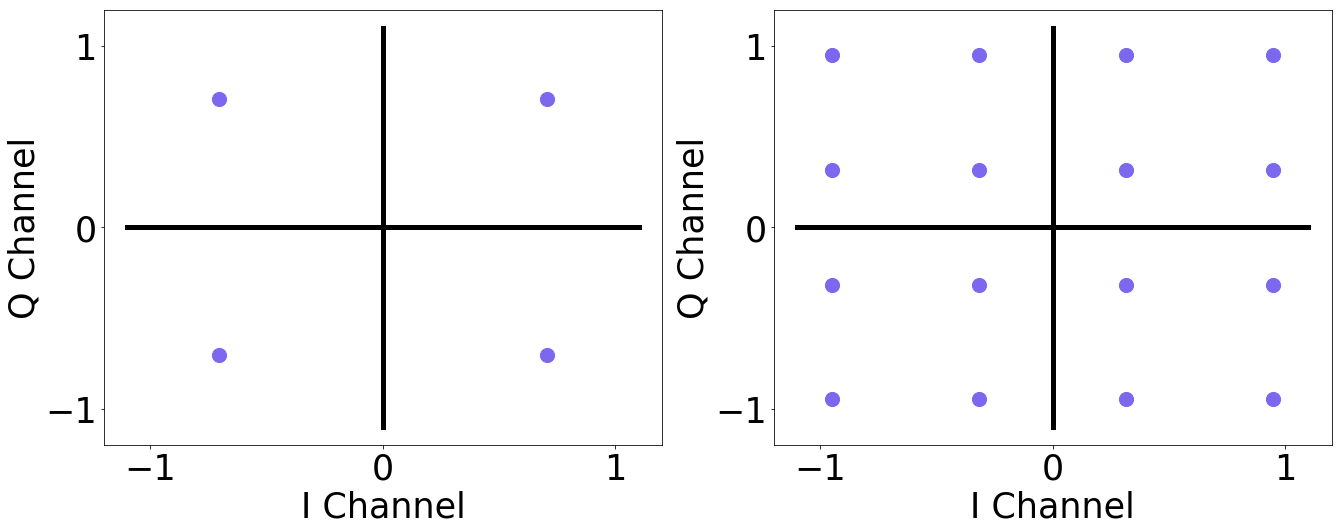}}
	\caption{Constellation points under the power constraint. Left: The QPSK modulation scheme. Right: The 16-QAM modulation scheme.}
	\label{Conspoint}
\end{figure}

In the following simulations, we consider two classical modulation schemes: QPSK and 16-QAM. Under the power constraint \eqref{pc}, their constellation points (also known as the alphabet) are illustrated in Fig \ref{Conspoint}. For both cases, all the constellation points in $\mcX$ are restricted in the region $[-1,1]\times[-1,1]$.

%To be more specific, the constellation points of the QPSK and the 16-QAM modulation schemes under the power constraint \eqref{pc} is illustrated in Fig \ref{Conspoint},
%
%where the constellation points (also known as the alphabet) $\mcX$ is bounded in $[-1,1]\times[-1,1]$ in these two cases.

Meanwhile, according to the channel law \eqref{awgneq}, the channel output \textcolor{black}{alphabet} $\mcY$ \textcolor{black}{is} $\mbbR^2$ itself. Under some reasonable assumptions, e.g. $\eta\le 1,\sigma_z^2\le \frac{1}{2}$, we are able to truncate the \textcolor{black}{alphabet} $\mcY$ with a \textcolor{black}{sufficiently large region, e.g.} $[-8, 8]\times[-8, 8]$. Correspondingly, we can discretize the \textcolor{black}{region} by a set of uniform grid points $\{y_j\}_{j=1}^{N}$:
\begin{equation*}
\begin{aligned}
    y_{r\sqrt{N}+s} &= (-8 + r\Delta y, -8 + (s-1)\Delta y), \Delta y = \frac{16}{\sqrt{N}-1},\\
    r &= 0,1,\cdots,\sqrt{N}-1, s = 1,2,\cdots,\sqrt{N},
\end{aligned}
\end{equation*}

%
% three-sigma rule

%Meanwhile, due to the AWGN channel rule \eqref{awgneq}, the received signal $Y$ equals the transmit signal $X$ plus a Gaussian white noise $Z \sim N(0,\sigma_{z}^{2})$. 
%
%Considering the the three-sigma principle of the Gaussian distribution, 
%
%we are able to truncate the region $\mcY$ as
%
%
%\begin{equation}\label{3_sig}
%	\bigcup_{i=1}^{M}\{\ y\ |\ \|y - x_i\|_{2} < 3\sigma_{z}\} \subseteq [-8, 8]\times[-8, 8]. 
%\end{equation}
%
%In particular, here $\{y_j\}_{j=1}^{N}$ are set as the uniform grid points
%\begin{equation}
%\begin{aligned}
%    y_{r\sqrt{N}+s} &= (-8 + r\Delta y, -8 + (s-1)\Delta y), \Delta y = \frac{16}{\sqrt{N}-1},\\
%    r &= 0,1,\cdots,\sqrt{N}-1, s = 1,2,\cdots,\sqrt{N},
%\end{aligned}
%\end{equation}
%on the interval $[-8, 8]\times[-8,8]$. 
%
In Algorithm \ref{alg:OT}, the termination condition of the Newton iteration is that the update step size of $\lambda$ is less than $10^{-14}$. Moreover, the $\text{SNR} \triangleq 1/(2\sigma_{Z}^{2})$ is used in the sequel. 

%The Newton iteration for solving $G(\lambda) = 0$ in the proposed algorithm will terminate when the difference between the $\lambda$ in the two most recent iterations is less than the convergence tolerance $\textit{tol}=10^{-14}$. 
%
%Note that $\text{SNR} \triangleq 1/(2\sigma_{Z}^{2})$ is used in the sequel. 
%

All the experiments are conducted on a platform with 128G RAM, and one Intel(R) Xeon(R) Gold 5117 CPU @2.00GHz with 14 cores.
%
%
%Remind that $\text{SNR} \triangleq \frac{1}{2\sigma_{Z}^{2}}$

\subsection{Algorithm Verfication}
We first study the convergence of our Sinkhorn algorithm by considering the residual errors of \eqref{sink_iter} and \eqref{G_lambda}:
%First, we would like to clarify the convergence behavior of our proposed Sinkhorn algorithm by considering the residual errors of \eqref{sink_iter} and \eqref{G_lambda} with respect to the $L_{1}$ norm:
%
\begin{equation*}
\begin{aligned}
r_{\phi} &= \sum_{i=1}^{M}|\phi_{i}\sum_{j=1}^{N}\Lambda_{ij}\psi_{j} - \mu_{i}|,\\
r_{\psi} &= \sum_{j=1}^{N}|\psi_{j}\sum_{i=1}^{M}\Lambda_{ij}\phi_{i} - \nu_{j}|,\\
r_{\lambda} &= |G(\lambda)|.
\end{aligned}
\end{equation*}
The parameters are set to
\begin{equation} \label{expe01}
    (\eta,\theta) = (0.9, \pi/18), \; N = 250,000, \;
    \text{SNR} = 0 \textrm{dB}.
\end{equation}
\begin{figure}[H]
	\centerline{\includegraphics[width=0.5\textwidth]{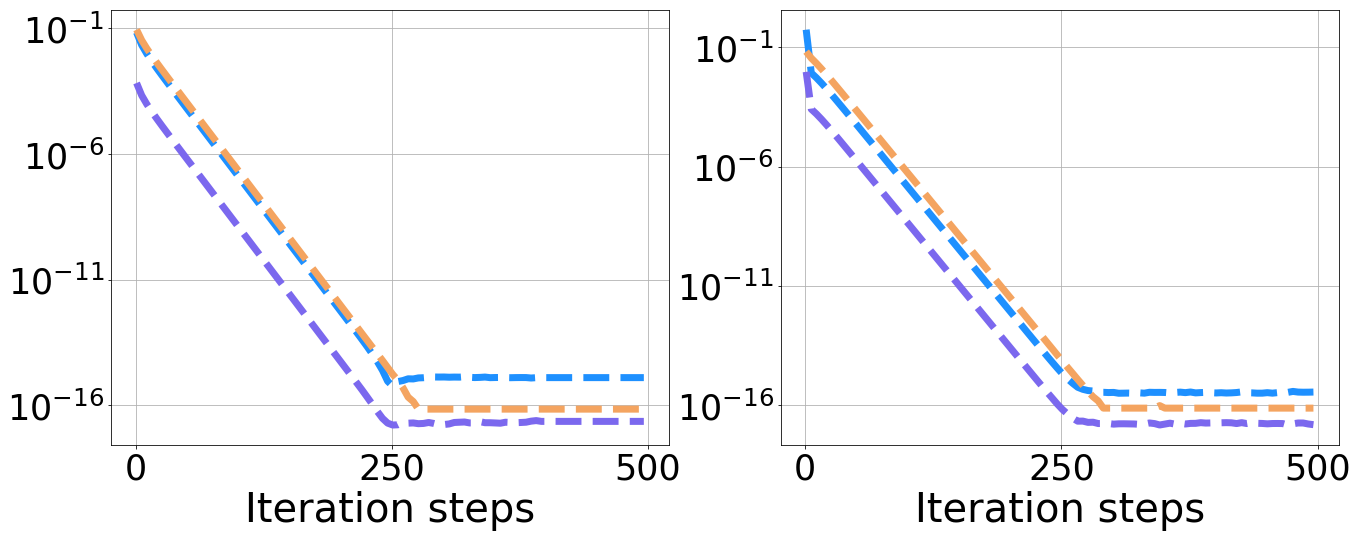}}
	\caption{The convergent trajectories of the residual error for $r_{\phi}$ (Orange), $r_{\psi}$ (Blue) and $r_{\lambda}$ (Purple). Left: The QPSK modulation scheme. Right: The 16-QAM modulation scheme.}
	\label{Res}
\end{figure}
%The convergence behavior of $r_{\phi}, r_{\psi}, r_{\lambda}$ in each iteration is plotted in Fig. \ref{Res} and we can clearly find that the three lines of $r_{\phi}, r_{\psi}, r_{\lambda}$ will all decline to less than the $\textit{tol}=10^{-14}$ rapidly.
%
%In fact, it needs only hundreds of iterations to reach such a high accuracy.
%
%Thus, we can claim that the optimal solution given by our proposed algorithm meets all the restrictions with high accuracy and stability.
%
%It is also worth mentioning that the traditional Sinkhorn algorithm usually takes tens of thousands of iterations to converge, while the convergence of our proposed algorithm is reached after only hundreds of iterations, which further implies the effectiveness of our proposed algorithm.

\begin{table*}[ht]\label{compare} %%参数： h:放在此处 t:放在顶端 b:放在底端 p:在本页
	\renewcommand\arraystretch{1.25}
	\centering  % 显示位置为中间
	\caption{Comparison between the Sinkhorn algorithm and CVX. Columns 3-5 are the averaged computational time and the speed-up ratio of the Sinkhorn algorithm. Column 6 is the averaged difference of the LM rate computed by two methods.}  % 表格标题
	\label{Table_cvx}  % 用于索引表格的标签
	%字母的个数对应列数，|代表分割线
	% l代表左对齐，c代表居中，r代表右对齐
	%\textbf{Table I}\\  %%表的标题
	\setlength{\tabcolsep}{6.5mm}{
		\begin{tabular}{c|c|c|c|c|c} %第一列设置宽度为45pt 全为左对齐 没有分割线
			%\setlength{\tabcolsep}{20mm}
			%\hline  % 表格的横线
			\toprule % 顶部线
			\multirow{2}{*}{} & \multirow{2}{*}{$N$} & \multicolumn{2}{c|}{Computational time (s)} &
			\multirow{2}{*}{Speed-up ratio} & \multirow{2}{*}{Average difference}\\
			\cline{3-4}
			 &  & Sinkhorn & CVX & &\\%[3pt]只改一行    %%表格第一行标题 %% 表格中的内容，用&分开，\\表示下一行
			\hline  % 表格的横线
			%\midrule % 中部线
			\multirow{3}{*}{QPSK} & $100$ & $0.37\times10^{0}$ & $2.61\times10^{1}$ & $7.05\times10^{1}$ & $1.99\times10^{-7}$\\
			 & $225$ & $0.64\times10^{0}$ & $8.66\times10^{1}$ & $1.35\times10^{2}$ & $6.45\times10^{-7}$\\
			 & $400$ & $1.28\times10^{0}$ & - & - & -\\
			%$4$ & $1600$ & $2.93\times10^{0}$ & - & - & -\\
			\hline
			%%%%%%%%%%%%%%%%%%%%%%%%%%%%
			\multirow{3}{*}{16-QAM} & $100$ & $0.96\times10^{0}$ & $1.49\times10^{2}$ & $1.55\times10^{2}$ & $4.13\times10^{-7}$\\
			 & $225$ & $1.96\times10^{0}$ & $9.27\times10^{2}$ & $4.73\times10^{2}$ & $1.37\times10^{-6}$\\
			 & $400$ & $3.81\times10^{0}$ & - & - & -\\
			%$16$ & $1600$ & $1.01\times10^{1}$ & - & - & -\\
			\hline
			\multirow{2}{*}{256-QAM} & $100$ & $1.49\times10^{1}$ & - & - & -\\
			%$256$ & $225$ & $2.32\times10^{1}$ & - & - & -\\
			%$256$ & $400$ & $4.95\times10^{1}$ & - & - & -\\
			 & $1600$ & $2.24\times10^{2}$ & - & - & -\\
			\bottomrule % 底部线
			\multicolumn{6}{}{}\\[1pt]
			%\hline  % 表格的横线
			%\multicolumn{cols}{pos}{text}
			%\multirow{number of rows}{width}{text}
	\end{tabular}}
\end{table*}

In Fig. \ref{Res}, we output the convergent trajectories of the residual errors with respect to iteration steps. We can see that the three curves all decrease rapidly and reach the machine accuracy near $250$ iterations. It is worth mentioning that classical OT problems usually require tens of thousands of Sinkhorn iterations to converge. This significantly illustrates the \textcolor{black}{dramatic} efficiency advantage of our OT model and Sinkhorn algorithm. 

To further illustrate the accuracy and efficiency of the Sinkhorn algorithm for the OT problem \eqref{OT}. We use CVX \cite{2010cvx} as the baseline. The averaged computational time and the averaged difference of the optimal values between the two methods are \textcolor{black}{listed} in Table \ref{Table_cvx}. To reduce the influence of noise, we repeat each experiment for $100$ times. Except for $N$, other parameters are the same as \eqref{expe01}. \textcolor{black}{Since it is difficult for CVX to handle large-scale problems, we restrict $N$ to small scales, e.g. $100,\;225,\;400$ and $1600$.} From the table, we can see the optimal values obtained by the two methods are almost the same. But our Sinkhorn algorithm has a significant advantage (one or two orders of magnitude) in computational speed. Moreover, for slightly larger scale problems, CVX has failed to output convergent results. 

\subsection{Results and Discussions}

Below, we present the computational results of the LM rate under different modulation schemes, different parameters $(\eta,\theta)$, and different SNRs. For comparison, we also present the computational results of GMI \cite{zhang2011general} under the same setup. As we know, GMI is \textcolor{black}{generally lower} than the LM rate \cite{2020Information}. These results not only help us quantitatively understand the relationship between the two rates, but also help verify the correctness of our model and algorithm. In the numerical \textcolor{black}{experiments}, we consider four sets of parameters $(\eta,\theta)$, namely,
%In this section, we compare the LM rate calculated by the Sinkhorn algorithm against the GMI calculated using the method of \cite{zhang2011general}. Since GMI should be no larger than the LM rate \cite{bai2020information}, we choose GMI as a benchmark. The QPSK and the 16-QAM modulation schemes are discussed respectively, and we select the following mismatch cases to carry out the experiments:
\begin{equation*}
   (0.9, \frac{\pi}{18}),\quad
   (0.9, \frac{\pi}{12}),\quad
   (0.8, \frac{\pi}{18}), \quad
   (0.8, \frac{\pi}{12}).
\end{equation*}
We set $N=250,000$ to \textcolor{black}{ensure} the discretization accuracy. And we use $500$ iterations for each \textcolor{black}{experiment} to ensure the algorithm \textcolor{black}{convergence}. We also repeat each \textcolor{black}{experiment} for $100$ times to reduce the influence of noise.
\begin{figure}[H]
	\centerline{\includegraphics[width=0.40\textwidth]{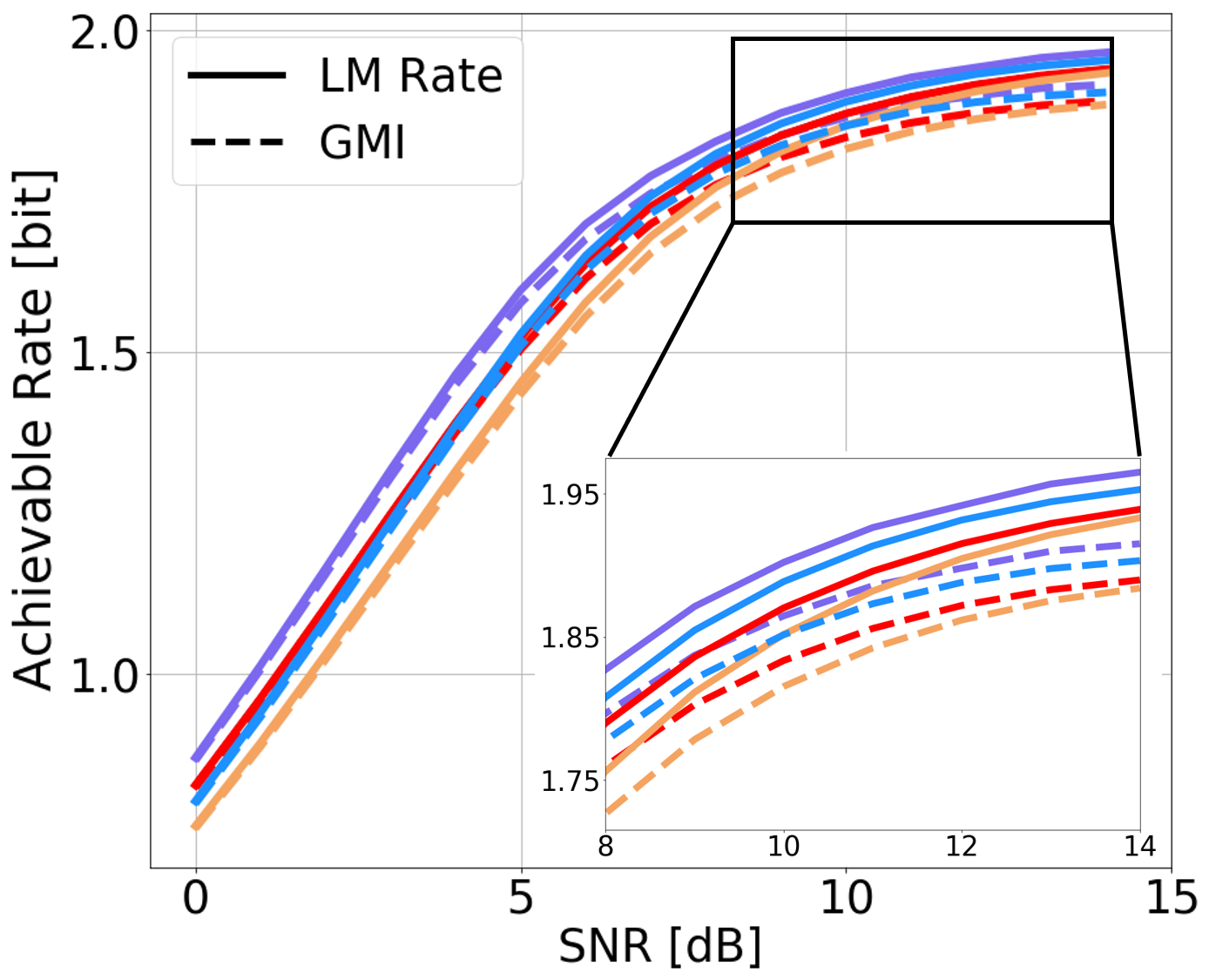}}
	\caption{LM rate (solid) and GMI (dashed) versus SNR for the QPSK modulation scheme under different mismatched cases, including $(\eta,\theta) = (0.9, \pi/18)$ (Purple), $(\eta,\theta) = (0.8, \pi/18)$ (Blue), $(\eta,\theta) = (0.9, \pi/12)$ (Red), and $(\eta,\theta) = (0.8, \pi/12)$ (Orange).}
	\label{QPSK}
\end{figure}

\begin{figure}[H]
	\centerline{\includegraphics[width=0.40\textwidth]{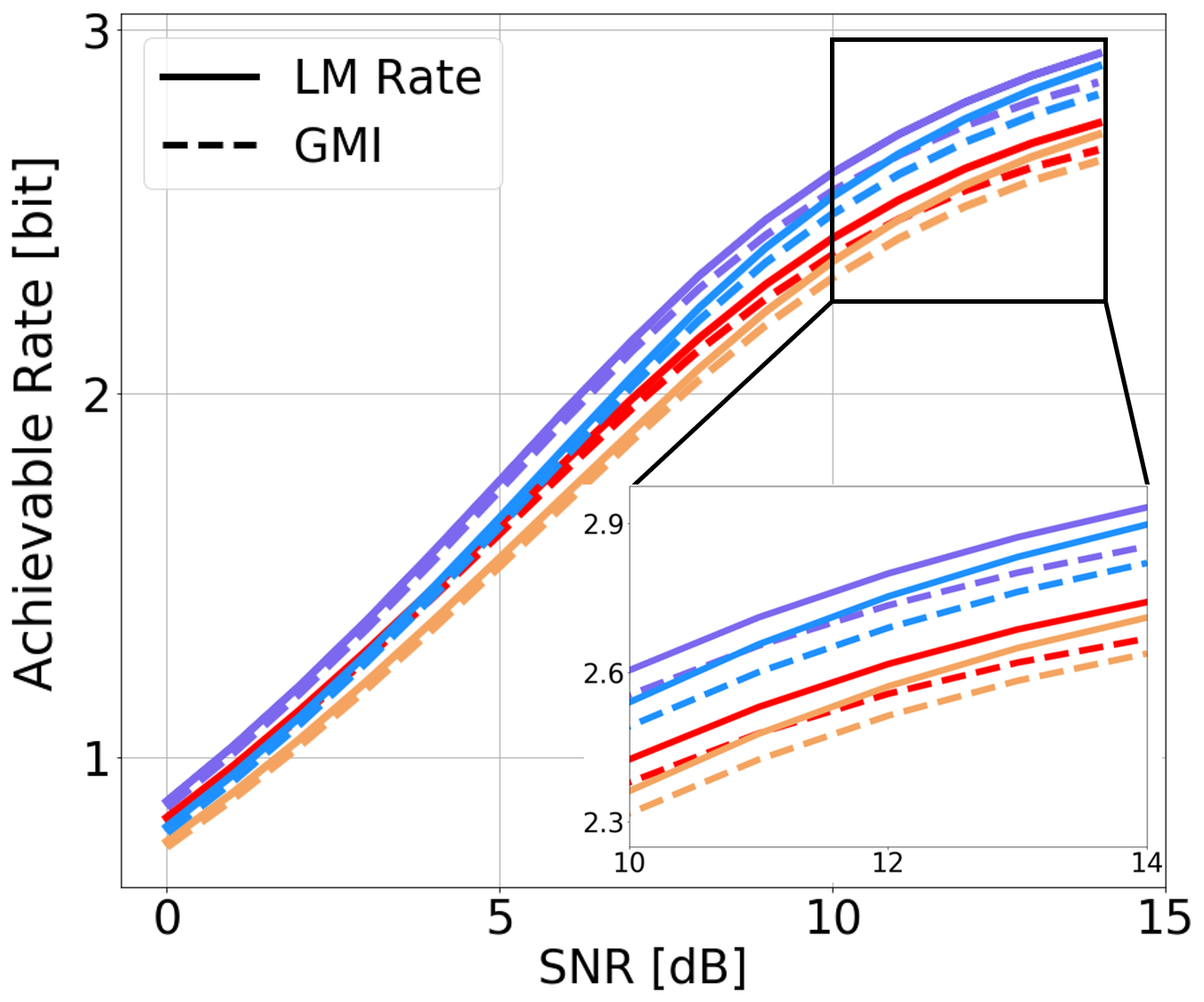}}
	\caption{LM rate (solid) and GMI (dashed) versus SNR for the 16-QAM modulation scheme under different mismatched cases, including $(\eta,\theta) = (0.9, \pi/18)$ (Purple), $(\eta,\theta) = (0.8, \pi/18)$ (Blue), $(\eta,\theta) = (0.9, \pi/12)$ (Red), and $(\eta,\theta) = (0.8, \pi/12)$ (Orange).}
	\label{16QAM}
\end{figure}
In Fig. \ref{QPSK}, we can see the comparison of the LM rate and GMI verse SNR for the QPSK modulation scheme. We observe that the LM rate is \textcolor{black}{higher} than GMI with the same parameters. Especially when $\text{SNR} > 8$dB, there is about $2\%$ gain in the LM rate. In \textcolor{black}{addition}, as $\eta$ decreases (from $0.9$ to $0.8$) or $\theta$ increases (from $\pi/18$ to $\pi/12$), both LM rate and GMI decrease accordingly. These results agree with intuition. In Fig. \ref{16QAM}, we \textcolor{black}{display} the comparison for the 16-QAM modulation scheme. From this, we can draw the same conclusions as those for Fig. \ref{QPSK}. 

\section{CONCLUSION} \label{sec_conclu_5}

In this paper, we \textcolor{black}{studied} the computation problem of the LM rate, which is a lower bound for mismatch capacity. Our contributions are twofold. First, we showed that the computation of the LM rate \textcolor{black}{can} be reformulated into the Optimal Transport problem with an extra constraint. Second, we proposed a Sinkhorn-type algorithm to solve the above problem. For the extra constraint, we show that it is equivalent to seeking the root of a one-dimensional monotonic \textcolor{black}{function}. Numerical \textcolor{black}{experiments} show that our approach to computing the LM rate is efficient and accurate. Moreover, we can observe a \textcolor{black}{noticeable} gain in the LM rate compared to GMI.

\bibliographystyle{bibliography/IEEEtran}
\bibliography{bibliography/LM_REF}
%\bibliography{references}  

\end{document}